 \documentclass[final,3p,times,twocolumn]{elsarticle}
\usepackage{graphicx}
\usepackage[cmex10]{amsmath}

\usepackage{wrapfig}

\usepackage{lineno}
\linenumbers

\journal{Nuclear Instruments and Methods in Physics Research A}

\begin{document}

\begin{frontmatter}



\title{A Study of Reactor Neutrino Monitoring \\ at Experimental Fast Reactor JOYO}


\author{H. Furuta$^{a,1,*}$, Y. Fukuda$^b$, T. Hara$^c$, T. Haruna$^{d, 2}$, N. Ishihara$^e$, M. Ishitsuka$^a$, C. Ito$^f$, M. Katsumata$^g$, T. Kawasaki$^g$, T. Konno$^a$,  M. Kuze$^a$, J. Maeda$^{a, 3}$, T. Matsubara$^a$, H.Miyata$^g$, Y. Nagasaka$^h$, K. Nitta$^{a, 4}$, Y. Sakamoto$^i$, F. Suekane$^j$, T. Sumiyoshi$^d$, H. Tabata$^j$, M. Takamatsu$^f$, N. Tamura$^g$}

\address{$^a$ Department of Physics, Tokyo Institute of Technology, Tokyo 152-8551, Japan
                 $^b$ Department of Physics, Miyagi University of Education, Sendai 980-0845, Japan
                 $^c$ Department of Physics, Kobe University, Kobe 657-8501, Japan
                 $^d$ Department of Physics, Tokyo Metropolitan University, Hachioji 192-0397, Japan
                 $^e$ Institute of Particle and Nuclear Studies, High Energy Accelerator Research Organization (KEK), Tsukuba 305-0801, Japan
                 $^f$Experimental Fast Reactor Department, Oarai Research and Development Center, Japan Atomic Energy Agency (JAEA), Oarai, 311-1393, Japan
                 $^g$ Department of Physics, Niigata University, Niigata 950-2181, Japan
                 $^h$ Department of Computer Science, Hiroshima Institute of Technology, Hiroshima 731-5193, Japan
                 $^i$ Department of Information Science, Tohoku Gakuin University, Sendai 981-3193, Japan
                 $^j$ Department of Physics, Tohoku University, Sendai 980-8578, Japan}

\begin{abstract}
We carried out a study of neutrino detection at the experimental fast reactor JOYO using a 0.76\,tons gadolinium loaded liquid scintillator detector.
The detector was set up on the ground level at 24.3\,m from the JOYO reactor core of 140\,MW thermal power.
The measured neutrino event rate from reactor on-off comparison was 1.11$\pm$1.24(stat.)$\pm$0.46(syst.)\,events/day. Although the statistical significance of the measurement was not enough, the background in such a compact detector at the ground level was studied in detail and MC simulation was found to describe the data well. A study for improvement of the detector for future such experiments is also shown.

\end{abstract}


%\footnote{$^{*}$ Corresponding Author. Tel.: +81 22 795 6727.\\
%                  Email address: furuta@awa.tohoku.ac.jp (H. Furuta).\\
%                 $^{1}$ Present Address: Department of Physics, Tohoku University, Sendai 980-8578, Japan}

\end{frontmatter}


\section{Introduction}
\label{Introduction}
\renewcommand{\thefootnote}{\fnsymbol{footnote}}
\footnotetext[1]{ Corresponding Author. Tel.: +81 22 795 6727.

Email address: furuta@awa.tohoku.ac.jp (H. Furuta).}
\renewcommand{\thefootnote}{\arabic{footnote}}
\footnotetext[1]{Present Address: Department of Physics, Tohoku University, Sendai 980-8578, Japan}
\footnotetext[2]{Present Address: Canon Inc., Tokyo 146-8501, Japan}
\footnotetext[3]{Present Address: Department of Physics, Tokyo Metropolitan University, Hachioji 192-0397, Japan}
\footnotetext[4]{Present Address: National Institute of Radiological Sciences, Chiba 263-8555, Japan}
Reactor neutrinos have been playing an important role since its first discovery in 1956~\cite{Reines} for the progress of elementary particle physics and to deepen our understanding of the nature.  
Now the reactor neutrino detection techniques have become mature after a number of reactor neutrino experiments so far performed~\cite{Reactor_exp}\cite{ABern}.
Research and development of compact reactor neutrino detector utilizing the up-to-date technologies have become active recently~\cite{AAP} with an idea of using it as a monitor for Plutonium breeding in reactor cores~\cite{ABern}\cite{MCribier} and as a very near detector to calibrate reactor neutrino flux for long baseline reactor neutrino oscillation experiments. 

\subsection{Reactor neutrinos}
\label{reactor-neutrino}
 
In operating reactors, $^{235}$U, $^{238}$U, $^{239}$Pu and $^{241}$Pu perform fission reaction after absorbing a neutron. 
The fission products are generally neutron-rich unstable nuclei and perform $\beta$-decays until they become stable nuclei. 
One $\bar{\nu}_e$ (anti-electron neutrino) is produced in each $\beta$-decay. 
The energy of the reactor neutrinos corresponds to $\beta$-decay energy of a few MeV. Roughly $6~\bar{\nu}_e's$ are produced in a fission reaction along with $\sim200$\,MeV of energy release, resulting in $6\times10^{20}~\bar{\nu}_e's$ production per second in a 3\,GW$_{th}$ power reactor.  

\subsection{Nondestructive Plutonium Measurement }
\label{Plutoniumu_measurement}
Main components of reactor neutrinos come from $^{235}$U and $^{239}$Pu fissions, and contributions of $^{238}$U and $^{241}$Pu are much smaller than those nuclei. 
 Along with the burn-up of the core, $^{235}$U is consumed and $^{239}$Pu is 'breeded' from $^{238}$U through neutron absorption and $\beta$-dacays.
Because $^{239}$Pu can be used for nuclear explosion, it is an important object of strict safeguard regulations.
Therefore, it is important to monitor reactor operation and track the plutonium breeding. 
International Atomic Energy Agency (IAEA) watches reactors in the world with surveillance cameras, reviewing operation record, etc. 
Because it is impossible to hide the neutrinos, it could be a powerful tool to monitor the reactor operation, in addition to the traditional monitoring methods~\cite{IAEA}.

The reactor neutrino monitoring has a potential to non-destructively measure the plutonium amount in the core.  

Table~\ref{nnuandle} shows the energy releases and expected number of emitted $\bar{\nu}_e$'s above 1.8\,MeV per fission, and average ratio of fission in the JOYO core for major isotopes in nuclear reactors.
As shown in the Table~\ref{nnuandle}, $^{235}$U produces significantly more neutrinos than $^{239}$Pu. 
Combining the neutrino flux and thermal power generation, there is a possibility to measure Plutonium amount in the core.  
This is simply depicted by the following equations assuming the fuel is made up only from $^{235}$U and $^{239}$Pu. 
\begin{eqnarray}
q_{235}F_{235}+q_{239}F_{239}=P_{th},\\
\nu_{235}F_{235}+\nu_{239}F_{239}=N_{\bar{\nu}_e}
\end{eqnarray}
where, 235 and 239 represent $^{235}$U and $^{239}$Pu. 
$F_x$ is the fission rate of the nucleus-$x$ in the core, $q_x$ is the energy release per fission. 
$\nu_x$ is the expected number of emitted $\bar{\nu}_e$'s per fission, $N_{\bar{\nu}_e}$ is the total emission rate of $\bar{\nu}_e$. 
A small contribution from $^{238}$U and $^{241}$Pu is ignored to simplify the calculation. 

The fission rate of $^{239}$Pu is calculated from those relations and the values of the parameters, and the $^{239}$Pu amount in the core can be calculated from the fission rate.

\begin{table}
\begin{center}
\small
\begin{tabular}{ccccc}
\hline
\hline
Isotope   &$\nu$ ($>$1.8\,MeV)  &q (MeV)& Contribution & \\
 & & & @JOYO(\%)\\\hline
$^{235}$U &1.92$\pm$0.02&201.7$\pm$0.6&  37.1 \\
$^{238}$U &2.38$\pm$0.02&205.0$\pm$0.9&  7.3\\
$^{239}$Pu&1.45$\pm$0.02&210.0$\pm$0.9&   51.3\\
$^{241}$Pu&1.83$\pm$0.02&212.4$\pm$1.0&   4.3\\
\hline
\hline
\end{tabular}
\caption{Number of $\bar{\nu}_e$ per fission with the energy above 1.8MeV~\cite{fissiondata} and energy release per fission for major isotopes in nuclear reactors~\cite{james}.}
\label{nnuandle}
\end{center}
\end{table}

\subsection{Compact neutrino detectors }
\label{Compact_detectors}

As R\&D of compact neutrino detectors, an experimental program led by Lawrence Livemore National Laboratory (LLNL) and Sandia National Laboratories (SNL) measured neutrino energy spectrum at a short distance from a $^{235}$U-rich reactor with a thermal power of 3.4\,GW$_{th}$, San Onofre Nuclear Generation Station (SONGS), and indicated feasibility of the neutrino monitoring~\cite{SONGS}.
On the other hand, further R\&D studies of detector design and materials are still necessary to realize a compact detector operation above ground for practical use as a reactor monitor with the neutrino detection.
Considering the neutrino interaction cross-section on proton target (inverse $\beta$-decay, $O(10^{-43})\,{\rm cm}^{2}$, see Section~\ref{neutrinodetect}) and compact detector size, the detector must be set at a short distance (less than a few tens of meters) from the reactor core to accumulate enough statistics for monitoring.
In addition, feasibility of the measurement at ground level is required for the monitor considering limited access to the reactor site, while the previous measurements of neutrinos were operated at underground to reduce cosmic-ray muon background.
Therefore, the detector must be designed to be able to reduce external backgrounds, e.g. cosmic-ray muons and fast neutrons.

We constructed a 0.76\,tons gadolinium loaded liquid scintillator detector as a prototype of KASKA detector~\cite{kaskaloi} and we reused it to take part in such R\&D efforts~\cite{furutaphd}. 
The detector was set up at 24.3\,m from Joyo experimental reactor core whose thermal energy was 140\,MW~\cite{joyopaper}. 
Unique points of this experiment are, 
(1) the reactor power is much smaller compared with the ones so far used to measure the neutrinos, 
(2) the detector is located above ground, 
(3) the reactor was a fast reactor, so that the neutrinos came mainly from Plutonium.
The main goal of this experiment was to distinguish reactor-on and off by neutrinos under this unfavorable conditions. 
One of the possible safeguard applications is to monitor small reactors to prevent them to be hiddenly operated to make plutonium. 
The points (1) and (2) of this experiment are useful to study such a possibility.  
As for (3), neutrinos from $^{235}$U-rich light water reactors have been measured~\cite{buggey}\cite{rovno}, while observation of neutrinos from $^{239}$Pu-rich fast reactor has not been reported yet and this experiment could have been the first detection of the fast reactor neutrinos. 
If energy spectrum of fast reactor neutrinos is measured in the future, 
$\nu_{235}$ and $\nu_{239}$ can be determined separately by comparing the $^{239}$Pu-rich neutrinos and $^{235}$U-rich neutrinos.
This experiment is a good practice to perform an experiment at a larger fast reactor in the future to measure the Plutonium-rich neutrino spectrum.

\section{Experimental fast reactor JOYO}
The experimental fast reactor JOYO, whose thermal power is 140\,MW, is located in Japan Atomic Energy Agency (JAEA) Oarai Research and Development Center in Ibaraki prefecture, Japan.
The JOYO reactor is a sodium-cooled fast reactor built as an experimental reactor to promote commercialization of fast breeder reactor development~\cite{joyopaper}.
The reactor fuel is plutonium-uranium mixed oxide (MOX) which consists of enriched uranium dioxide UO($_2$) to 18 w\% in $^{235}$U and plutonium dioxide PuO($_2$).
Fraction of fissile Pu content ($(^{239}Pu + ^{241}Pu)/all$) is about 16 w\% at the inner core and about 21 w\% at the outer core. 

\begin{figure}
\centering
\includegraphics[width=3.3in]{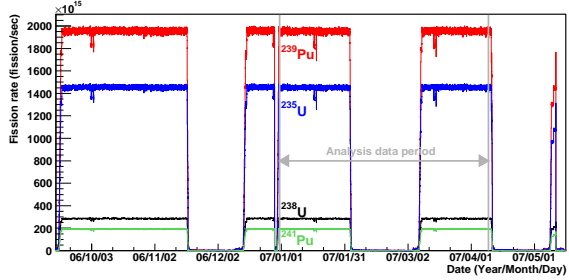}
\caption{Fission rate of each fissile element as a function of time from the 4th to 6th operational cycles of experimental fast reactor JOYO. 
Four lines correspond to $^{239}$Pu, $^{235}$U, $^{238}$U and $^{241}$Pu as indicated in the figure. A period used for data analysis is also shown.}
\label{joyofissionrate}
\end{figure}

JOYO reactor operates for 60 days then stops for a few weeks in its operational cycle.
Therefore, we could collect data in both the reactor-on and reactor-off conditions.
The data taken under reactor-off condition were used to measure the background.
Thermal power of the rector was stable at 140\,MW during its operation. 
Figure~\ref{joyofissionrate} shows time variations of fission rates of main isotopes ($^{235}$U, $^{238}$U, $^{239}$Pu and $^{241}$Pu) in the fuel.
Neutrino flux from the reactor core was calculated from available measurements of $\beta$-decay spectra with 2.5\,\% systematic uncertainty~\cite{fissiondata}.

\section{Neutrino detection principle}
\label{neutrinodetect}
\begin{figure}
\centering
\includegraphics[width=2.8in]{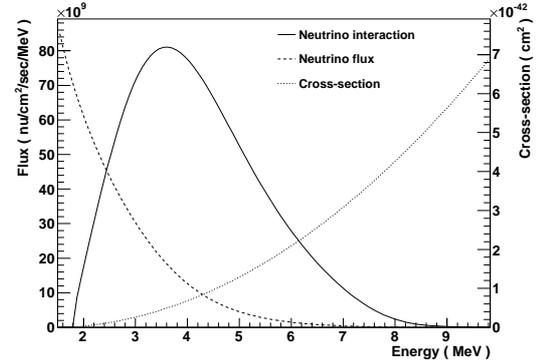}
\caption{Shape of neutrino energy spectrum (arbitrary unit) via inverse $\beta$-decay reactions expected in the detector (solid line). Overlaid curves show the reactor neutrino flux (dashed line) and cross-section of inverse $\beta$-decay reaction (dotted line).}
\label{nuespectra}
\end{figure}

Reactor neutrinos are detected with a liquid scintillator formulated from organic oils. 
Organic oils are abundant in free protons and the reactor $\bar{\nu}_e$ react with the proton through inverse $\beta$-decay reaction.
\begin{equation}
\bar{\nu}_e + p \rightarrow e^+ + n
\label{eq:ib_reaction}
\end{equation}
Figure~\ref{nuespectra} shows the reactor neutrino flux at JOYO experimental site and
the cross-section of inverse $\beta$-decay reaction together with a shape of
the energy spectrum in the detector.
Number of interactions in the detector is determined as a multiplication
of the flux, interaction cross-section and the number of free protons in
the detector.
Energy threshold of the inverse $\beta$-decay interaction is 1.8\,MeV.
Cross-section of inverse $\beta$-decay reaction is associated with the lifetime of free neutrons and calculated precisely with 0.2\,\% accuracy~\cite{Vogel}.

The detector contains gadolinium-loaded liquid scintillator (Gd-LS), in which neutrino signals are detected by using delayed coincidence technique.
A positron kinetic energy and $\gamma$'s from its annihilation are observed as the prompt signal.
Since the recoil energy of neutron is small, neutrino energy can be measured from the energy of the prompt signal.
\begin{equation}
E_{signal} = E_{\nu} - 1.8\,{\rm MeV} + 2m_{e}c^{2} 
\end{equation}
Neutrons from inverse $\beta$-decay reactions are captured by gadolinium or hydrogen in the Gd-LS mostly after thermalization, and $\gamma$-rays are emitted.
Those $\gamma$-rays are detected as delayed signal.
In our detector, we expect 76.9\,\% of neutron captures are on $^{155}$Gd or $^{157}$Gd, which have more than $10^{5}$ times larger thermal neutron capture cross-section than hydrogen~\cite{chooz}.
$\gamma$-rays with total energy of approximately 8\,MeV are emitted from a neutron capture on Gd.
The mean time difference ($\Delta t$) between the prompt and delayed signals is estimated to be 46\,$\mu$sec.
The background events are strongly suppressed by requiring coincidence of two signals.

\section{The detector}
\subsection{Experimental setup}
\begin{figure}
\centering
\includegraphics[width=3.0in]{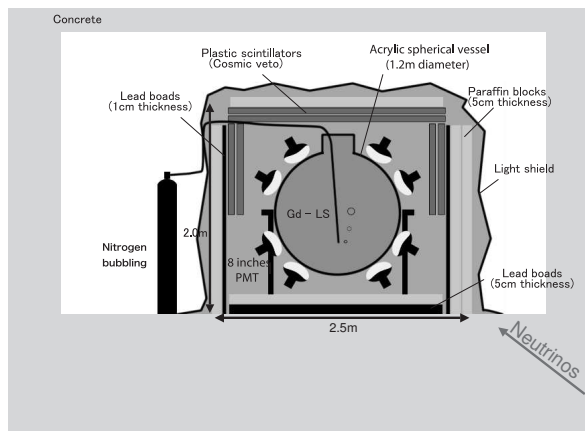}
\caption{Schematic view of experimental setup of the detector.}
\label{detectorall}
\end{figure}

\begin{figure}
\centering
\includegraphics[width=2.7in]{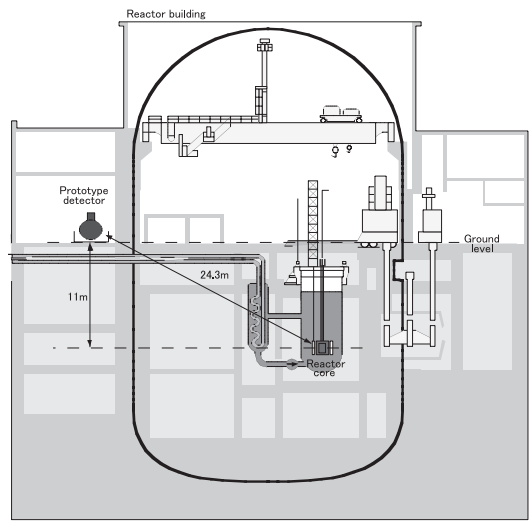}
\caption{Location of the detector at experimental fast reactor JOYO.}
\label{detectorsite}
\end{figure}

Figure~\ref{detectorall} shows a schematic view of the detector. 
The detector was constructed at Tohoku University as one of the R\&D programs for the KASKA reactor neutrino oscillation experiment~\cite{kaskaloi}.
The detector was moved to the Joyo reactor site in September 2006 after the KASKA R\&D studies, and was set up on the ground floor of the reactor building near a delivery entrance, just at the west outside of the reactor containment vessel.
The distance to the Joyo core was 24.3\,m.
The location of the detector is shown in Figure~\ref{detectorsite}.
Because Joyo uses sodium coolant, water was not allowed to be brought in the building and water shield was not possible. 

The data taking period was from January 2007 until December 2007.
Unfortunatelly the liquid scintillator deteriorated during the operation and only net 38.9 days reactor-on data and 18.5 days of reactor-off data were used for the analysis. 
The reason of the deterioration is not clear but we assume the high temperature environment and N$_2$ bubbling were possible reasons.  
Details of detector design and components are shown in the following sections.

\subsection{Main detector}
The main detector consisted of 0.76\,tons of Gd-LS filled in a transparent acrylic spherical vessel with inner diameter of 1.2\,m.
The vessel was made of UV transparent acrylic ACRYLITE(000) of MITSUBISHI RAYON Inc..
Two acrylic hemispheres were made from 15\,mm thick acrylic plates by vacuum forming.
The two hemispheres were put together sandwiching Viton O-ring at the equator to form a sphere. 
There is a 30\,cm diameter chimney at the top of the sphere. 
The acrylic sphere was supported by an aluminum stand which stood in a oil pan. 
The liquid scintillator was formulated by diluting the commercial Gd-loaded liquid scintillator BC521 (Saint-Gobain) by Paraffine oil and Pseudocumene. 
The compositions of the liquid scitillator were, 12.6\,weight\% (w\%) Pseudocumene (1,2,4-Trimethylbenzene: C$_{9}$H$_{12}$), 76.3\,w\% Paraol 850, 11.2\,w\% BC521, and 1.52\,g/liter of PPO (2,5-Diphenyloxazole: C$_{15}$H$_{11}$NO) as the fluor. The Gd concentration was 0.05\,w\%  (as contained in BC521). 
Paraol 850 is heavy isoparaffin, one of Shell products. 
The scintillation light yield was measured to be 56\,\% of Anthracene scintillator, which is equivalent to 9,400\,photons/MeV.
The Gd-LS was purged by N$_{2}$ bubbling with a flowing rate of 100\,cc/min during operation to reduce the oxygen quenching effect. 
Properties of Gd-LS used in our detector are summarized in Table~\ref{lstable}.
\begin{table}
\begin{center}
\small
\begin{tabular}{cc}
\hline
\hline
Parameter&Value \\ \hline
Density (20 $^{\circ}$C)&0.838 g/cm$^{3}$ \\
H/C ratio&1.94\\
Number of Protons~(H)& 6.22$\times$ 10$^{28}$\\
Light yield & 9,400\,photon/MeV\\
Gd concentration & 0.05\,w\% \\
Neutron capture time&46.4\,$\mu$sec \\ \hline
\hline
\end{tabular}
\caption{Properties of gadolinium-loaded liquid scintillator used in our detector}
\label{lstable}
\end{center}
\end{table}

\begin{figure}
\centering
\includegraphics[width=2.3in]{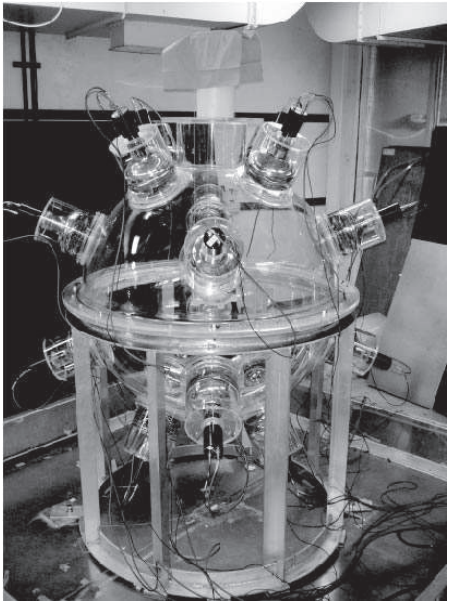}
\caption{Picture of sphere shape acrylic vessel equipped with 16 PMTs on the surface. Gd-LS is not filled at the time of this picture.}
\label{detectorpic}
\end{figure}

The scintillation lights from Gd-LS were measured by 16\,Hamamatsu R5912 8-inch photomultiplier tubes (PMTs) mounted on the surface of the acrylic vessel. 
Each PMT was covered by a mu-metal skirt which was used for Kamiokande PMT long time ago.  
Figure~\ref{detectorpic} shows a picture of the acrylic vessel with 16 PMTs on the surface. 
The PMT was put in a acrylic housing cylinder and the space between PMT surface and acrylic sphere were filled with RTV rubbers (Shin-Etsu Silicones KE103, KE1052). 
The photo-cathode coverage was approximately 10\,\% .

\subsection{Cosmic-ray veto counter and detector shielding}
\label{sec:shield}
\begin{figure}
\centering
\includegraphics[width=2.7in]{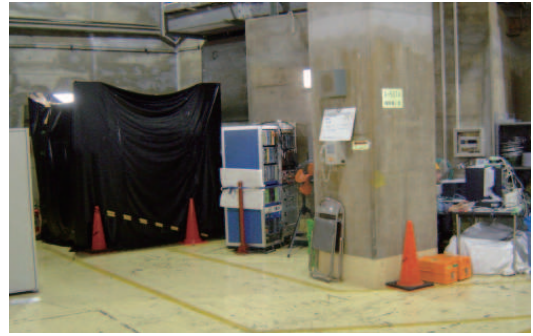}
\caption{A picture taken at the experimental site. The detector is in the black cottage at the left hand side. The reactor containment vessel is behind the concrete wall at the right hand side. }
\label{detector_site}
\end{figure}

Since the detector was set up at ground level, cosmic-ray muon flux was large.
In order to reduce the cosmic-ray muon background, the main detector was surrounded by a veto counter system.
This system consisted of two layers of 1\,cm thick plastic scintillator plates equipped with wavelength shifter and PMT for the readout.
Top of the detector and the north and south sides were fully covered by scintillator layers, while only the upper half were covered for the east and west sides.
Veto efficiency of cosmic-ray muons by the counter system was estimated to be 92\,\% from MC simulation including the acceptance.
The cosmic-ray veto signal rate was about 2\,kHz.
5\,cm thick lead blocks covered the bottom area of the oil pan and 6\,mm thick lead sheets backed by wooden boards covered the side of the detector housing. 
5\,cm thick paraffin blocks were arranged outside of the detector to suppress fast neutrons induced by cosmic muons.
Figure~\ref{detector_site} shows a picture of the detector at the site. 
The size of the detector was roughly 2.5\,m$\times$2.5\,m$\times$2\,m(H). 

\subsection{Data acquisition system}
\begin{figure}
\centering
\includegraphics[width=3.1in]{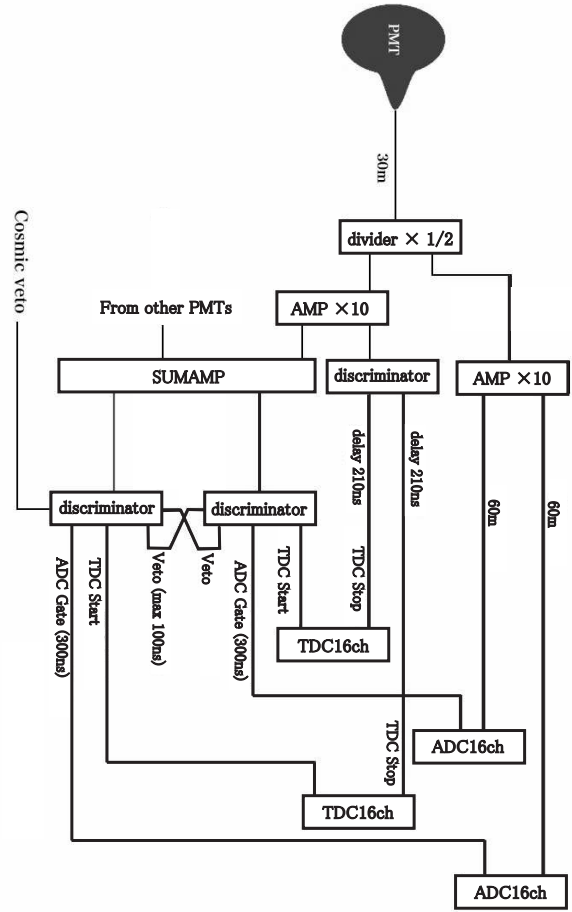}
\caption{The schematic view of the data acquisition system.}
\label{electronics}
\end{figure}

CAMAC standard electronics modules were used for data taking. Figure~\ref{electronics} shows the schematic view of the data acquisition system.
Signal from each PMT was divided into four.
The first one was fed into ADC, by which integrated charge within 200\,nsec time window was measured.
The second signal was used to make a common trigger for ADC and TDC, which was made from a discriminated analog sum of all PMT signals.
The threshold for the common trigger was set at 3.5\,MeV. If cosmic-ray veto counter had a signal within 100\,$\mu$sec before the common trigger was made, the trigger was canceled.
The third signal provided a stop signal to TDC, which measured the timing of the PMT hits.  
The last signal was fed into another ADC for pulse shape discrimination (PSD) study aiming to identify fast neutron background, although the PSD was not used for the study described in this paper.
In order to collect delayed coincidence signals from neutrino interactions, lower trigger threshold at 2.5\,MeV was applied to the delayed signals for 100\,$\mu$sec after a prompt trigger was created.
Time interval between the first and second triggers was measured by counting a 100\,MHz clock signal by a CAMAC scaler and the data were saved along with the ADC and TDC data for each trigger.
If a trigger for delayed signal was not generated within 100\,$\mu$sec, the data acquisition system was back to the normal mode with 3.5\,MeV threshold. In addition to this delayed trigger, we also took data with single trigger at 0.6\,MeV threshold for the background study.
During the data taking at JOYO fast reactor site, the single trigger rate was about 300\,Hz.
Mean dead time of the data taking inclusive of the cosmic muon veto time was 38\,\%. 

The readout and monitoring system in this experiment needed to be simplified due to limited access to the experimental area.
Therefore, we constructed our DAQ software system in a CAMAC CC/NET~\cite{sakamoto} to read the data from ADC and TDC modules and used a trigger system installaed into a NIM FPGA module.

As JOYO is a fast reactor and uses sodium as moderator, the experimental area also needed to be kept off water and high humidity. In addition, the detector using liquid scintillator, which generates organic gas, was placed in a large box sealed with black vinyl sheets.
In order to keep safety of the experimental area during the operation of the experiment, we constructed a monitoring system in a Linux computer and kept watching temperature, humidity and density of oxygen and organic gases.
The experiment and monitoring data were automatically sent to a 220\,km distance remote site, Tohoku Gakuin University. We built a secure network on the internet by IPsec VPN architecture over IPv4 protocol, which enabled an experiment shift person to check the condition of DAQ and experiment area remotely~\cite{sakamoto2}.

\subsection{Monte Carlo simulation}
\begin{figure}
\centering
\includegraphics[width=3.0in]{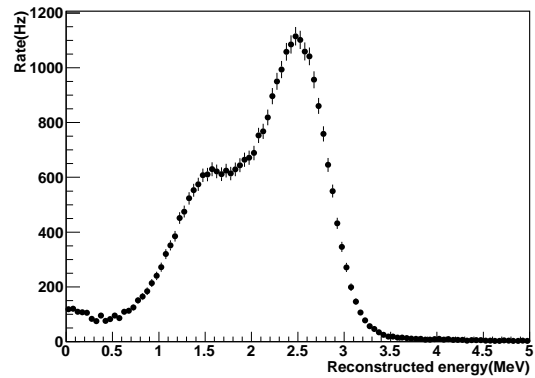}
\caption{Reconstructed energy spectrum from data taken with $^{60}Co$ $\gamma$-ray source at the detector center.}
\label{energycalbrationE}
\end{figure}

The collected data were compared with the Monte Carlo (MC) simulation based on Geant4 (version 4.9.0.p1).
Geant4 is a toolkit which provides a calculation of particle tracking in materials~\cite{geant4}.
For the hadronic interaction process, QGSP\_BIC\_HP model~\cite{hadp} was employed in Geant4.
It comprehends from low energy region under 20\,MeV such as behavior of thermal and fast neutron to high energy region such as interactions between cosmic-ray muons and materials around it.
Trajectory of optical photons emitted in the Gd-LS was simulated considering the optical process including attenuation and scattering.

Corrections for PMT responses and energy calibration were carried out by putting a $^{60}$Co $\gamma$-ray source inside the detector.
Energy was reconstructed from the total observed charge by 16 PMTs in which correction to the acceptance and attenuation length in the liquid scintillator were taken into account.
Figure~\ref{energycalbrationE} shows a reconstructed energy spectrum from the data taken with a $^{60}$Co $\gamma$-ray source at the detector center. The $^{60}$Co source mainly emits two gamma rays with 1.17\,MeV and 1.33\,MeV energies. A large peak in Figure~\ref{energycalbrationE} is made from the gamma rays with 2.5\,MeV total energy.
The energy resolution estimated from the peak at 2.5\,MeV was 20\,\%/$\sqrt{E({\rm MeV})}$.

In addition, the measured data with $^{241}$Am-$^{9}$Be ($\alpha$, n) neutron source at the detector center were used to tune the quenching effects of protons recoiled by neutrons parametrized by Birks' constant $k_{B}$~\cite{birks} and evaluate the neutrino MC simulation.
The Birks' constant of our Gd-LS was estimated to be 0.07\,mm/MeV from a comparison of the measured energy spectrum to the MC simulation.

Not only the neutrino signal events, but also various background events were generated by the MC simulation and compared with the observed data.
Those background events included cosmic-ray muons and the muon decay, fast neutrons and environmental $\gamma$-rays from decay chains of $^{238}$U and $^{232}$Th series and  $^{40}$K decays.
In addition to the fast neutron and environmental $\gamma$-rays generated inside the detector, those from outside of the detector were also considered in the MC simulation.

\section{Measurement of background spectrum}

\begin{figure}
\centering
\includegraphics[width=3.0in]{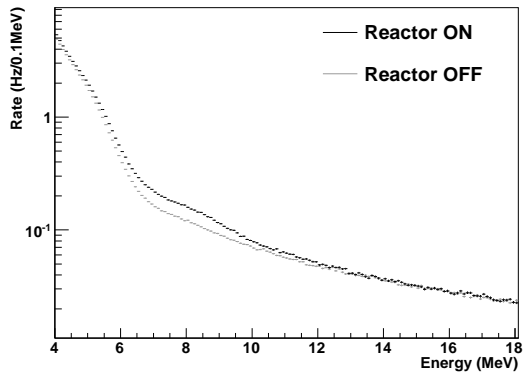}
\caption{Comparison between reconstructed energy spectra of the prompt trigger events for the reactor-on and off. Black and gray histograms show the observed data for a day live-time under reactor-on and off conditions, respectively. There is an excess around 8\,MeV of the distributions attributed to thermal neutron capture on Gd. }
\label{thermalex}
\end{figure}

\begin{figure*}
\centering
\includegraphics[width=2.8in]{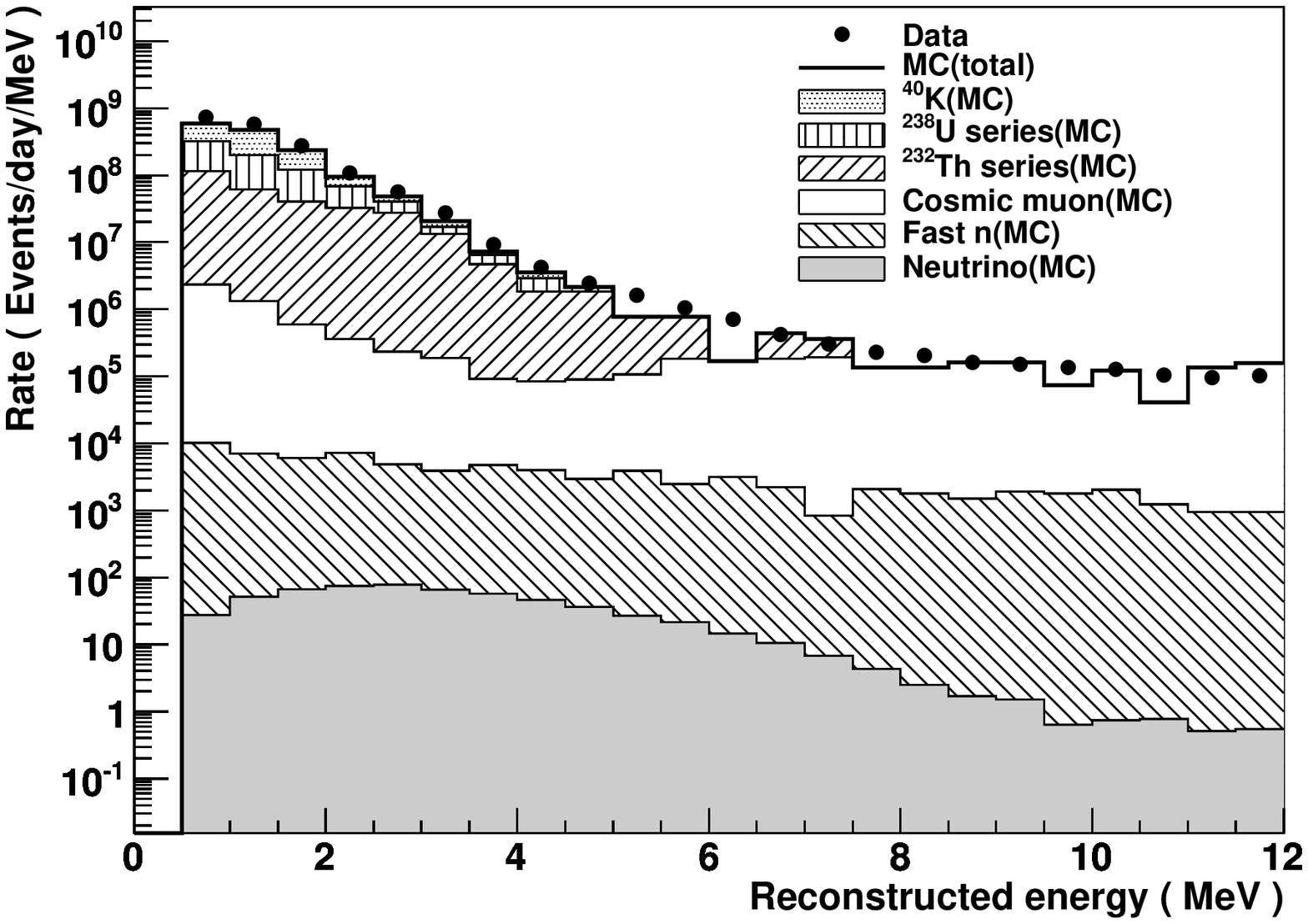}
\includegraphics[width=2.8in]{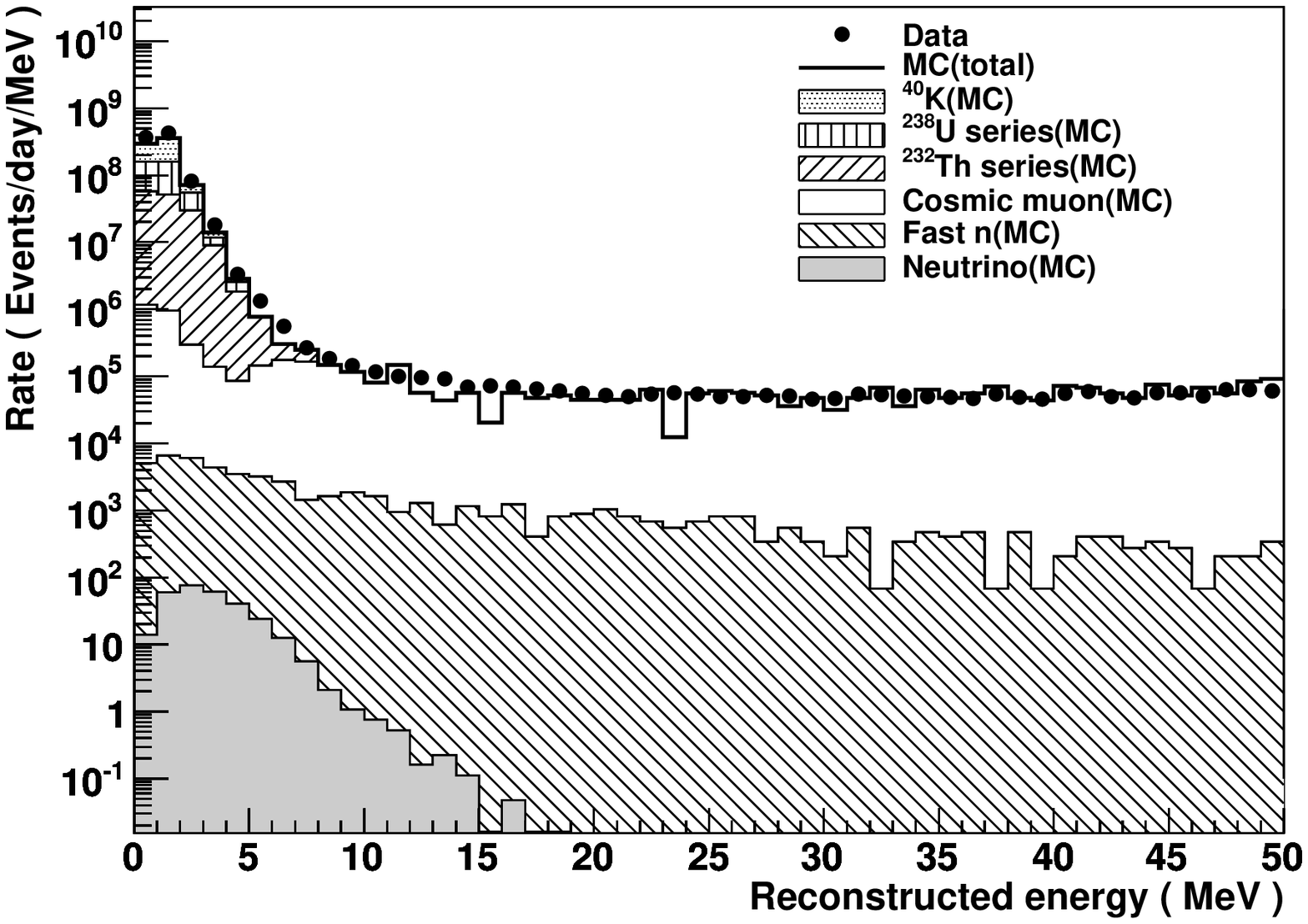}
\caption{Reconstructed energy spectra of the prompt trigger events above 0.6MeV of the threshold level with different energy ranges.  
Points show the observed data for 309\,sec live-time taken in reactor-off condition. 
Overlaid histograms show the expected neutrino signal and background energy spectrum with the contributions from each background source.}
\label{singlebg}
\end{figure*}

Major background sources in this experiment were environmental $\gamma$-rays and cosmic-ray muons.
The environmental $\gamma$-rays are emitted by radioactive isotopes contaminated in the detector and materials around the detector.
These $\gamma$-rays are produced through the decay chains of $^{238}$U and $^{232}$Th series, and decay of $^{40}$K.
The energy of $\gamma$-rays ranges up to 2.6\,MeV.
However, there were $\gamma$-ray contaminations above the discriminator threshold level of 3.5\,MeV due to the energy resolution tail. 
The main source of the $\gamma$-rays was considered to be concrete walls surrounding the detector. 
Cosmic-ray muons have wide energy range over GeV scale.
High energy muons generate fast neutrons by interactions in materials composing the experimental site and fast neutrons turned out to be the severest background for neutrino signals in this experiment. 
Most of the background events produced by muons were excluded by the delayed coincidence technique but there were still remaining backgrounds even after requiring it.
Those background events were further reduced by the data analysis as explained in later sections.

Huge number of neutrons were produced in the reactor core, and a very small fraction of them could reach the neutrino detector passing through materials constructing the Joyo or crevices in the materials. Low energy neutrons are detected as gamma-rays emitted via neutron capture on Gd. Figure~\ref{thermalex} shows a comparison of  the reconstructed energy spectra for the reactor-on and off. An excess in reactor-on was found around 8\,MeV, which was considered to be made by the low energy neutrons from the core. The excess rate between 4\,MeV to 14\,MeV was 6.1\,Hz. Because energy distribution of the thermal neutron was unknown, it was impossible to precisely calculate the detection efficiency. Therefore we carried out rough estimation of the thermal neutrino flux. Assuming the detection efficiency of 8\,MeV gamma rays was 50\% and considering naively the total cross section for the neutrons to the detector was equivalent to the surface area of the acrylic sphere, the neutron flux at the detector site can be estimated to be approximately 10$^{-4}$/cm$^{2}$/sec. This neutron flux is too small to be detected by usual neutron counters.

The background events which satisfy the delayed coincidence condition were classified into two categories, namely accidental and correlated backgrounds. 
The accidental background consists of two independent background events which accidentally occur within the delayed coincidence time window.
The main source of such background events are environmental $\gamma$-rays followed by cosmic-ray muons.
The correlated backgrounds are caused by a continuous physics process.
Those processes include decays of cosmic-ray muons inside the detector and fast neutrons induced by cosmic-ray muon followed by neutron capture on Gd in the detector. 
In the former case, cosmic-ray muon causes background to the prompt signal, and an electron from the muon decay (Michel electron) is identified as the delayed signal.
In the latter case, recoil protons caused by a fast neutron are detected as a prompt signal, and the $\gamma$-rays from neutron capture on Gd are identified as a delayed signal.
Some of radioactive isotopes produced by cosmic-ray muon interactions in the detector cause coincidence signals in its decay chain and can be considered as background source~\cite{KamLAND-cosmo}.
However, the production rate of such isotopes was negligibly small compared to the other background sources in this experiment. 

Figure~\ref{singlebg} shows the reconstructed energy spectrum of the observed data with 0.6\.MeV of the threshold level together with the expected reactor neutrino signal and background events from the MC simulation, in which delayed coincidence cut condition was not required yet.
Simulation of cosmic-ray muon background was based on flux measured in \cite{bess}, and correction factor 0.72 was applied from a fit to the data for the high energy region between 20\,MeV and 140\,MeV.
Fast neutron flux was obtained as 0.63\,neutrons/cm$^2$/sec above 10\,MeV (equivalent to visible electron energy of 3\,MeV in the detector) from a fit to the distribution of time interval ($\Delta t$) between the prompt and delayed signals.
As is shown in Figure~\ref{singlebg}, environmental $\gamma$-ray background is dominant for the energy below 6\,MeV before the delayed coincidence condition is applied.
We assumed typical concentration of radioactive isotopes in concrete materials around the detector, 2.1\,ppm of $^{238}$U, 5.1\,ppm of $^{232}$Th and 1.4\,ppm of $^{40}$K~\cite{furutanim}.
The observed energy spectrum was in reasonable agreement with the expected background spectrum from the MC simulation after normalization corrections were applied to the cosmic-ray muon and fast neutron flux.
As is shown in this figure, the background level is $10^{5} \sim 10^{7}$ times higher than the neutrino signals before the delayed coincidence.
In order to measure reactor neutrino signals significantly over the background, reduction of the background events by the delayed coincidence technique and further selections are necessary.

\section{Neutrino event selection and results}
\label{sec:selection}
\begin{figure*}
\centering
\includegraphics[width=4.9in]{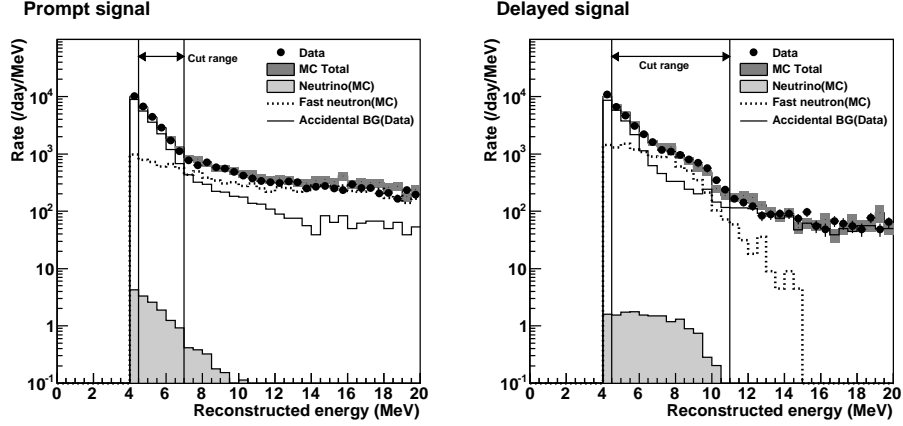}
\caption{Reconstructed energy spectra for the prompt (left) and delayed (right) signal candidate events. Points show the observed data for 7.4 hours live-time under reactor-off condition. Overlaid histograms show the expected spectra of neutrino signals and background events from MC simulation. }
\label{energynocut}
\end{figure*}

We used the data taken with delayed coincidence trigger to search for neutrino events.
In order to reduce the backgrounds, following selection criteria were applied to the data and the MC simulation.
\begin{itemize}
\item{$4.5 \le E_{prompt} \le 7\,{\rm MeV}$ and\\
          $4.5 \le E_{delayed} \le 11\,{\rm MeV}$},
\end{itemize}
where $E_{prompt}$ and $E_{delayed}$ are the reconstructed energy for the prompt and delayed signals, respectively.
Figure~\ref{energynocut} shows the reconstructed energy spectra for the prompt and delayed signals.
The lower cut value at 4.5\,MeV was set to reject environmental $\gamma$-rays, while the higher cut values were set at 7\,MeV and 11\,MeV to select prompt signal shown in Figure~\ref{nuespectra} and total 8\,MeV $\gamma$-rays from neutron capture on Gd, respectively. Especially, the delayed energy selection is very effective for rejection of Michel electron events with 53MeV of end point, in contrast to the fast neutron events induced by cosmic muons which have a similar distribution to neutrino delayed signals.

\begin{figure}
\centering
\includegraphics[width=2.5in]{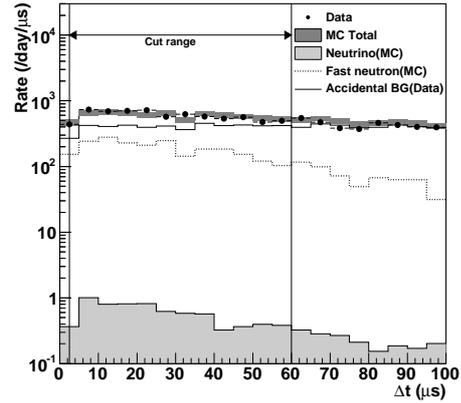}
\caption{$\Delta t$ distributions for events which satisfy the energy cut: $4.5 \le E_{prompt}\le 7$\,MeV and $4.5 \le E_{delayed} \le 11$\,MeV. Points show the observed data for 7.4 hours live-time taken in reactor-off condition. Overlaid histograms show the expectations of neutrino signal and background events from MC simulation.}
\label{dtnocut}
\end{figure}

\begin{itemize}
\item{$2.5 \le \Delta t \le 60\,\mu$sec}
\end{itemize}
Figure~\ref{dtnocut} shows the time difference between the prompt and delayed signals ($\Delta t$) for the events after the energy cuts.
Michel electron events have distribution following 2.2\,$\mu$s of muon life time, while the fast neutron and neutrino events have corresponding distributions with the decay time of 46.4\,$\mu$s, which is determined by Gd concentration in liquid scintillator.
Although the coincidence condition within 100\,$\mu$s time window was required in the data acquisition, further cut was applied in the analysis. The lower limit was used to reject remaining Michel electron events after the energy cuts, while the upper limit was set to collect enough neutrino events with 46.4\,$\mu$s decay time.

\begin{figure*}
\centering
\includegraphics[width=4.9in]{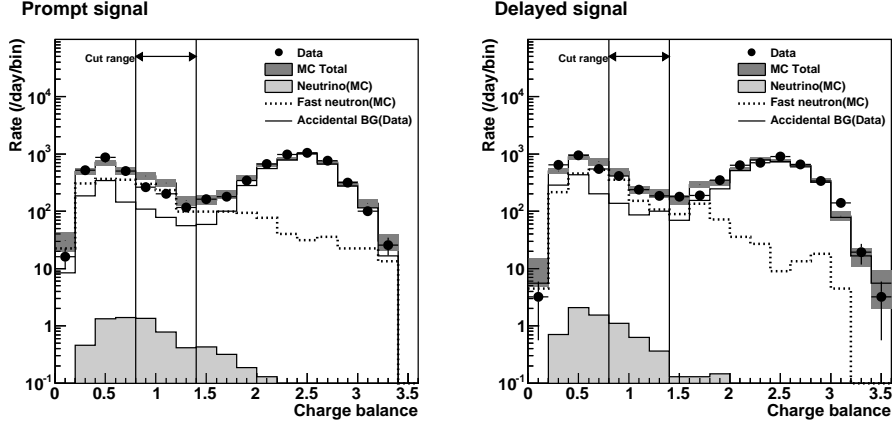}
\caption{Charge balance ($CB$) distributions for the events which satisfy energy and $\Delta t$ cuts.
Left-hand and right-hand figures show the distributions for prompt and delayed signals, respectively.
Points show the observed data for 7.4 hours live-time taken in reactor-off condition. Overlaid histograms show expected distributions of neutrino signal and background events from MC simulation.}
\label{cbnocut}
\end{figure*}

In order to further reduce background events remained after the energy and $\Delta t$ cuts were applied, charge balance ($CB$) is defined as follows:
\begin{equation}
\label{cbequation2}
CB=\sqrt{\dfrac{16 \left(\sum^{16}_{i=1}{ \left(Q^{cor}_{i}\right)^{2}}\right)}{\left( \sum^{16}_{i=1}{Q^{cor}_{i}}\right)^{2}}-1},
\end{equation}
where $Q^{corr}_{i}$ is observed charge from $i$-th PMT after gain correction was applied. 
This variable becomes large for the external background events, such as environmental gamma rays and fast neutrons, with the vertex position close to the surface of the detector while it is smaller for events occuring near the center. So $CB$ cut corresponds to a kind of fiducial volume cut.
However when the vertex position is too close to the surface and between the near PMT surfaces, the $CB$ becomes smaller and the vertex position mimics a place around the center of the detector, because solid angles to the near PMT surfaces from the vertex position are narrow and number of photoelectrons for the near PMTs becomes less.    
Figure~\ref{cbnocut} shows the $CB$ distributions of reactor neutrino and background events after energy and $\Delta t$ cuts were applied.
Distributions for the background events have a valley at $CB \sim 1.3$.
To maximize signal over noise ratio (S/N), cut conditions for $CB$ were defined as:
\begin{itemize}
\item{$0.8 \le CB_{prompt} \le 1.4$ and $0.8 \le CB_{delayed} \le 1.4$}
\end{itemize}

\begin{table}
\begin{center}
\small
\begin{tabular}{ccc}
\hline
\hline
Selection Criterion & Event rate (/day) & S/N ratio\\ \hline
Trigger level  & 162 & -\\
Energy cut & 8.89 & 1/1197\\
Coincidence cut & 7.01 & 1/1009\\
Charge balance cut & 0.988 & 1/128\\
Vertex $\phi$ cut & 0.494 & 1/34.6\\ \hline
\hline
\end{tabular}
\caption{Effect of selection criteria on the reactor neutrino events in the detector. Each cut condition is described in Section~\ref{sec:selection}.}
\label{cutefficiency}
\end{center}
\end{table}

\begin{figure*}
\centering
\includegraphics[width=4.9in]{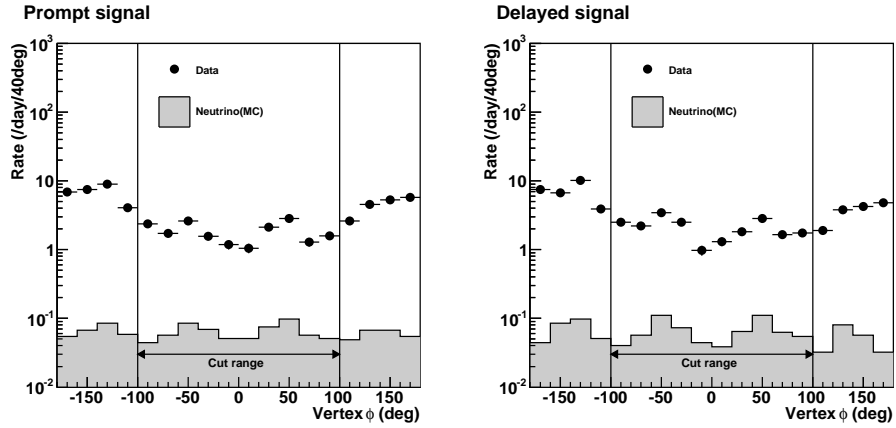}
\caption{Reconstructed vertex $\phi$ distribution after the cuts of energy, $\Delta t$ and $CB$ were applied. Points shows the data for 18.5 days live-time taken in reactor-off condition. An overlaid histogram shows the expectation of neutrino signals.} 
\label{vposcut}
\end{figure*}

Even after the energy, $\Delta t$ and charge balance cuts were applied, the remaining background events were still hundred times larger than the reactor neutrino events. 
The dominant component of the remaining background events was fast neutrons generated from muons in materials around the detector. 
Figure~\ref{vposcut} shows the vertex $\phi$ distributions for the prompt and delayed signals after energy, $\Delta$t and charge balance cuts were applied.
$\phi$ is the azimuthal angle in spherical polar coordinates as the $z$-axis vertical to the ground. Then $\phi=0$ was defined as north of the detector.
The vertex position is reconstructed by a fit with expected charge of each PMT based on the scintillation light yield and a solid angle to the PMT from the vertex position. 
It is expected that the vertex positions of the neutrino interactions distribute uniformly in the detector.
On the other hand, the vertex $\phi$ distributions of the data are not flat due to asymmetric arrangement of building materials and paraffin shields.
Therefore, we applied the following cut based on the vertex $\phi$ position to maximize S/N ratio:
\begin{itemize}
\item{$ -100^{\circ} \le \phi_{prompt} \le 100^{\circ}$ and \\
$-100^{\circ} \le \phi_{delayed} \le 100^{\circ}$}
\end{itemize}

The neutrino event selection cuts and expected event rates are summarized in Table~\ref{cutefficiency}.

\subsection{Result of the reactor neutrino event selection}
\begin{figure}[t]
\centering
\includegraphics[width=2.9in]{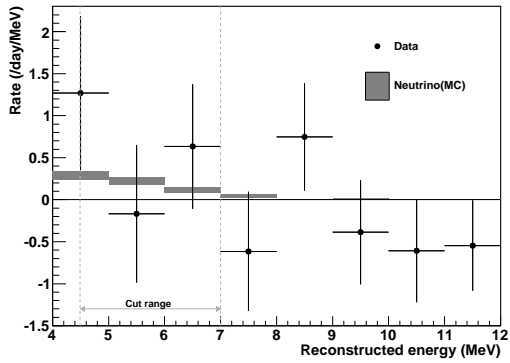}
\caption{Reconstructed energy spectrum after neutrino event selection criteria except for the prompt energy cut was applied. Points show the data with the statistical errors in which energy spectrum measured for reactor-off is subtracted from that taken for reactor-on. Accidental background event rates were estimeted from the data and subtracted. Boxes show the expected reactor neutrino energy spectrum from MC simulation with the MC statistical errors.}
\label{selectionplot}
\end{figure}

The result of the neutrino event selection is summarized in Table~\ref{resultselection}.
The accidental background event rates were estimated from the measurements of single background event rates and subtracted from the total event rates as shown in Table~\ref{resultselection}.
Then, the difference between the event rates for reactor-on and reactor-off was calculated.
Errors in Table~\ref{resultselection} are only the statistical ones.
After reactor neutrino event selection and subtraction of accidental background and reactor-off data, event rate for the neutrino candidate events from 38.9 days of reactor-on data and 18.5 days of rector-off data was obtained as 1.11$\pm$1.24(stat.)$\pm$0.46(syst.)\,events/day, while the expected neutrino signal event rate from the MC simulation was 0.49\,events/day.
The systematic uncertainties were estimated considering the uncertainties in energy resolution, energy scale and vertex reconstruction.
Figure~\ref{selectionplot} shows the prompt energy spectrum after all selections except for the prompt energy cut.
The measured excess rate was consistent with the expected neutrino rate from the MC simulation, but also consistent with zero within the systematic error. So the observation of neutrinos from experimental fast reactor JOYO has not been statistically established in this measurement. 
A new design of the detector is described in the following section, in which sensitivity to the measurement of reactor neutrinos was estimated based on the observed background event rates shown in this paper.

\begin{table}
\begin{center}
\small
\begin{tabular}{ccccc}
\hline
\hline
 & Reactor-on & Reactor-off & $\Delta$(on $-$ off) \\ 
Live-time & 38.9 days & 18.5 days & - \\
Total & 19.0$\pm$0.7 & 17.1$\pm$1.0 & 1.93$\pm$1.19\\
Accidental & 2.34$\pm$0.24 & 1.52$\pm$0.29 & 0.82$\pm$0.38\\
Correlated & 16.7$\pm$0.7 & 15.6$\pm$1.0 & 1.11$\pm$1.24\\\hline
Reactor $\nu$ MC & - & - & 0.494\\ \hline
\hline
\end{tabular}
\caption{Observed event rates (events/day) and the statistical uncertainties after the neutrino event selection criteria were applied. Accidental background event rates were estimated by single background event rate. Correlated event rates were obtained by subtracting the accidental BG event rate from the total event rate.}
\label{resultselection}
\end{center}
\end{table}

\section{New detector design for the next experiment}
\label{newdesign}
Problems found in the JOYO experiment were following.
\begin{itemize}
\item[(1)] {Statistics of the data was limited because the long term data taking was impossible due to degradation of Gd-LS.}
\item[(2)] {Energy threshold level could not be sufficiently lowered due to large amount of environmental $\gamma$-rays from outside of the detector. Therefore, the neutrino detection efficiency was obliged to be low.}
\item[(3)] {Performance of event vertex reconstruction was not enough to distinguish external background events entering the detector due to the same problem caused by the detector structure as in the description for $CB$ in Section~\ref{sec:selection}.}
\item[(4)] {Fast neutron background level was too high against the neutrino signals, for which S/N ratio was 0.029\,.}
\end{itemize}
We designed a new detector for the next experiment solving these problems, and estimated the sensitivity to reactor neutrino measurements based on the MC simulation.

We are considering a liquid scintillator with high Pseudocumene concentration above 99\,w\% as possible candidate to solve the Gd-LS degradation problem.
The high aromatic concentration is supposed to stabilize the Gd-LS.
Experimental studies of the long-term stability and characteristics of the proposed Pseudocumene based Gd-LS are necessary.
In addition, the detector design needs to be improved to suppress fast neutron background.
The new detector will consist of two concentric sphere vessels.
The inner vessel contains the Gd-LS as target of reactor neutrinos.
The outer vessel is filled with paraffin oil with no scintillation light emission, and works as shield against fast neutrons. In the MC simulation, we assumed the same target volume with Gd-LS, surrounded by 20\,cm layer of buffer oil.
Environmental $\gamma$-rays are also reduced by the outer layer, by which we estimate that the energy threshold can be lowered to 3\,MeV with the same trigger rate as the measurement at JOYO. 
The scintillation lights are viewed by 24 10-inch PMTs isotropically arranged on the surface of the outer vessel, providing 11\,\% photo-cathode coverage close to JOYO detector.
The PMT surfaces are away from the target vessel with the interval buffer region by which performance of vertex position reconstruction can be improved especially for those close to the surface of the target vessel. The reconstructed vertex radii in the polar coordinate system are used for rejection of the external background events. 

According to the study using the MC simulation, we expect the S/N ratio is improved from 0.029 to 0.093, with about 16 times larger neutrino selection efficiency by the new detector design.
If we put the same detector at MONJU rector site~\cite{monju}, of which thermal power is approximately five times higher than JOYO reactor, the S/N ratio is further improved to 0.48 with 41\,events/day of neutrino observations.
Significance of fast reactor neutrino observation reaches 2 standard deviation after 12\,days of reactor-on and off live-times at JOYO reactor site or 1\,day at MONJU reactor site.

\section{Conclusions}
We carried out an experimental study of fast reactor neutrino detection and measured the background spectrum at fast reactor JOYO using a compact detector.
The observed reactor neutrino candidate signal was 1.11$\pm$1.24(stat.)$\pm$0.46(syst.)\,events/day after subtraction of background events while the expected reactor neutrino event rate from the MC simulation was 0.49\,events/day.
As a result, the first observation of fast reactor neutrinos was not statistically established in this measurement.
On the other hand, various background sources at the ground level nearby the reactor were studied in detail and those backgrounds were found to be reproduced well by our MC simulation.
These background studies will be useful not only for the R\&D of future reactor neutrino oscillation experiments but also for the development of compact reactor neutrino detector as a remote monitor for plutonium breeding.

A design concept of a new detector and its sensitivity to the observation of fast reactor neutrinos were also described in this paper based on the knowledge acquired by the measurement at JOYO.
We expect the S/N ratio will be improved from 0.029 to 0.093 by the new detector design, and it is further improved to 0.48 if we put the detector at MONJU reactor site at the same distance from the core.
Expected reactor neutrino signal by the new detector design is 8.0\,events/day and 41\,events/day at JOYO and MONJU reactor site, respectively.

\section{Acknowledgements}
This work was supported by a grant-in-aid for scientific research (\#16204015) of Japan Ministry of Education, Culture, Sports, Science and Technology (MEXT).
This work was performed in cooperation with Japan Atomic Energy Agency (JAEA). Especially, we would like to thank T. Aoyama and T. Kuroha for supporting us in various ways. 
We thank K2K experimental group for providing the muon veto counters for this measurement.

\end{document}